\documentclass[fleqn,usenatbib,useAMS]{mnras}

\usepackage[T1]{fontenc}
\usepackage{ae,aecompl}

\usepackage{graphicx}	% Including figure files
\usepackage{amsmath}	% Advanced maths commands
\usepackage{amssymb}	% Extra maths symbols
\usepackage{multicol}   % Multi-column entries in tables
\usepackage{bm}         % Bold maths symbols, including upright Greek
\usepackage{pdflscape}  % Landscape pages

\title[The early evolution of clusters near the GC]{The early dynamical evolution of star clusters near the Galactic Centre}

\author[S.-M.~Park, S.~P.~Goodwin and S.~S.~Kim]{
So-Myoung Park,$^{1}$\thanks{E-mail: smpark12@khu.ac.kr}
Simon P. Goodwin$^{2}$
and Sungsoo S. Kim$^{1,3}$
\\
$^{1}$School of Space Research, Kyung Hee University, 1732, Deogyeong-daero, Giheung-gu, Yongin-si, Gyeonggi-do 17104, South Korea\\
$^{2}$Department of Physics and Astronomy, University of Sheffield, Sheffield S3 7RH, UK\\
$^{3}$Department of Astronomy and Space Science, Kyung Hee University, 1732, Deogyeong-daero, Giheung-gu, Yongin-si, Gyeonggi-do 17104, South Korea
}

\date{Accepted XXX. Received YYY; in original form ZZZ}

\pubyear{2018}

\begin{document}
\label{firstpage}
\pagerange{\pageref{firstpage}--\pageref{lastpage}}
\maketitle

\begin{abstract}
We examine the dynamical evolution of both Plummer sphere and substructured (fractal) star forming regions in Galactic Centre (GC) strong tidal fields to see what initial conditions could give rise to an Arches-like massive star cluster by $\sim 2$~Myr.  We find that any initial distribution has to be contained within its initial tidal radius to survive, which sets a lower limit of the initial density of the Arches of $\sim$ 600~M$_\odot$~pc$^{-3}$ if the Arches is at 30~pc from the GC, or $\sim$ 200~M$_\odot$~pc$^{-3}$ if the Arches is at 100~pc from the GC.  Plummer spheres that survive change little other than to dynamically mass segregate, but initially fractal distributions rapidly erase substructure, dynamically mass segregate and by 2~Myr look extremely similar to initial Plummer spheres, therefore it is almost impossible to determine the initial conditions of clusters in strong tidal fields. 

\end{abstract}

\begin{keywords}
methods: numerical - stars: formation -- Stars: kinematics and dynamics -  Galaxy: centre, Galaxy: open clusters and associations: individual: the Arches cluster 
\end{keywords}

%%%%%%%%%%%%%%%%%%%%%%%%%%%%%%%%%%%%%%%%%%%%%%%%%%

%%%%%%%%%%%%%%%%% BODY OF PAPER %%%%%%%%%%%%%%%%%%
\numberwithin{equation}{section}
\section{Introduction}

When we observe star clusters, we almost always see that they are relaxed, dense, spherical objects \citep{de Grijs et al.2002a,de Grijs et al.2002b,de Grijs et al.2002c,Lada&Lada2003,Gouliermis et al.2004,Sana et al.2010, Pang et al.2013}, but star forming regions are clumpy, filamentary and substructured \citep{Elmegreen2002,Cartwright&Whitworth2004,Zinnecker&Yorke2007,McKee&Ostriker2007,Schneider et al.2012,Konyves et al.2015}. This has led to a possible picture of star cluster formation as the merger/relaxation of substructure, however simulations of this have so far been without tidal fields which is a reasonable approximation to the Solar Neighbourhood \citep{Allison et al.2009a,Allison et al.2010,Parker et al.2011,Parker et al.2014}.

The Arches and Quintuplet clusters \citep{Figer et al.1999a,Figer et al.1999b,Figer et al.2002,Najarro et al.2004} are examples of young massive star clusters we observe near the Galactic Centre (GC) where the tidal field is extremely strong. In particular, the Arches cluster is a well-studied young (2-4 Myr; \citealt{Najarro et al.2004,Martins et al.2008}), massive ($\sim 2 \times 10^{4}$ M$_\odot$; \citealt{Kim et al.2000,Clarkson et al.2012}) and mass segregated \citep{Figer et al.1999b,Stolte et al.2002,Kim et al.2006,Espinoza et al.2009,Habibi et al.2013,Hosek et al.2015} star cluster with a {\em projected} distance of $\sim$ 30~pc from the GC. 

In this paper we  investigate the possible initial conditions that could produce an Arches-like star cluster close to the GC in the presence of a very strong tidal field. We perform $N$-body simulations of both fractal and Plummer distributions with a range of initial sizes at a variety of distances from the GC. Our goal is to see what possible range of initial conditions could give rise to a cluster like the Arches, and which could not.

%%%%%%%%%%%%%%%%%%%%%%%%%%%%%%%%%%%%%%%%%%%%%%%%%%%%%%%%%%%%%%%%%%%%%
\section{Method}

We simulate the early dynamics of star clusters using Aarseth's {\sc nbody6} code \citep{Aarseth1999} with full (non-truncated) tidal forces \citep{Kim et al.2000}. We simulate a both smooth (Plummer sphere) and clumpy (fractal) initial conditions at distances of 30 and 100~pc from the GC. 

Star clusters are evolved for 2 Myr, the (minimum) age of the Arches cluster \citep{Najarro et al.2004,Martins et al.2008}; these simulations are computationally expensive which sets this fairly short timescale.

\subsection{Tidal forces}

The tidal radius is the distance from the centre of the star cluster where the gradient of the effective potential is locally zero. Therefore, outside the tidal radius, stars are influenced more by the external potential of the Galaxy than that of the cluster. Often, tidal fields are modelled simply by applying a cut-off at several tidal radii beyond which stars are `lost' (i.e. removed from the simulation). However, in a strong tidal field such as we are simulating, there is a significant tidal force within the tidal radius, therefore it is important to fully include the Galactic potential (i.e. the variation of the orbital angular velocity across the cluster as given by Eq. \ref{eq2.3} below cannot be ignored).

Note that throughout, when we refer to `tidal radius' we mean the tidal radius for a point mass.  In situations where a cluster extends over the `tidal radius' the mass interior to the `tidal radius' is less than we assume.

The Galactic potential is constructed using Oort's $A$ and $B$ constants \citep{Oort1927}:

\begin{equation}
	\label{eq2.1}
	\begin{aligned}
		& A(R_{\rm G})=-\frac{1}{2}R_{\rm G}\frac{\text{d}\Omega(R_{\rm G})}{\text{d}R_{\rm G}},\\
        & B(R_{\rm G})=-\Bigg\{\Omega+\frac{1}{2}R_{\rm G}\frac{\text{d}\Omega(R_{\rm G})}{\text{d}R_{\rm G}}\Bigg\},
	\end{aligned}
\end{equation}
where $R_{\rm G}$ is the Galactocentric distance, and $\Omega$ is the orbital angular velocity. To calculate these constants near the GC, \citet{Kim et al.2000} make two assumptions.

\begin{table}
	\centering
	\caption{A summary of initial conditions for the simulations at 30~pc from the GC. $R_{\rm t}$ is the nominal tidal radius, $R_{\rm c}$ is the total (outer) cluster radius, $R_{\rm h}$ is the half-mass radius, and $Q$ is the virial ratio.}
	\label{table1}
	\begin{tabular}{ccccccc}
		\hline 
        $\text{Model}$ &  $R_{\rm t}$ & $R_{\rm c}$ & $R_{\rm h}$ & $Q$ \\ 
        \hline 
        Fractal & $\sim$ 2.0~pc & $\sim$ 4.0~pc & $\sim$ 3.0~pc & 0.3\\
        &  & $\sim$ 2.0~pc & $\sim$ 1.0~pc & 0.5\\
        &  & $\sim$ 2.0~pc & $\sim$ 1.0~pc & 0.3\\
        &  & $\sim$ 1.0~pc & $\sim$ 0.7~pc & 0.5\\
        \hline
        Plummer & $\sim$ 2.0~pc & $\sim$ 4.0~pc & $\sim$ 3.0~pc & 0.5\\
        &  & $\sim$ 2.0~pc & $\sim$ 1.0~pc & 0.5\\
        & & $\sim$ 1.0~pc & $\sim$ 0.7~pc & 0.5\\
        \hline
	\end{tabular}
\end{table}

Firstly, the Galactic enclosed mass ($M_{\rm G}$) profile at $R_{\rm G}$ follows a power-law \citep{Kim et al.1999}:
\begin{equation}
	\label{eq2.2}
	\begin{aligned}
		& M_{\rm G}=5.5 \times 10^{7}\hspace{0.1 cm} \text{M}_{\odot} \Bigg(\frac{R_{\rm G}}{30 \hspace{0.1cm}\text{pc}} \Bigg)^{1.4},\\
        & M_{\rm G}=6.4 \times 10^{8}\hspace{0.1 cm} \text{M}_{\odot} \Bigg(\frac{R_{\rm G}}{100 \hspace{0.1cm}\text{pc}} \Bigg)^{1.6},
	\end{aligned}
\end{equation}
where $5.5 \times 10^{7}\hspace{0.1 cm} \text{M}_{\odot}$ and $6.4 \times 10^{8}\hspace{0.1 cm} \text{M}_{\odot}$ are the enclosed masses at 30 and 100~pc from the GC. We determine these enclosed masses from \citet[their Fig.~14]{Launhardt et al.2002}.

Secondly, the star clusters are moving in a circular orbit so that their orbital angular velocity is 
\begin{equation}
	\label{eq2.3}
	\Omega(R_{\rm G})=\frac{1}{R_{\rm G}}\sqrt{\frac{\text{G}M_{\rm G}(R_{\rm G})}{R_{\rm G}}},
\end{equation}
where G is a gravitational constant.
By using Eq. \ref{eq2.3}, we can obtain Oort's constants for Eq. \ref{eq2.1} 
\begin{equation}
\begin{aligned}
	& A=-\frac{\alpha-3}{4}\Omega(R_{\rm G}),\\
    & B=A-\Omega(R_{\rm G}),
\end{aligned}
\end{equation}
where $\alpha$ is the power-law index of Eq. \ref{eq2.2}. And the tidal radius, $R_{\rm t}$, of the star cluster is 
\begin{equation}
	\label{eq2.5}
	R_{\rm t}=\Bigg\{\frac{\text{G}M_{\rm c}}{4A(A-B)}\Bigg\}^{1/3},
\end{equation}
where $M_{\rm c}$ is the cluster total mass \citep{Aarseth1999}.

\begin{table}
	\centering
	\caption{A summary of initial conditions for the simulations at 100~pc from the GC. Where $R_{\rm t}$ is the tidal radius, $R_{\rm c}$ is the total (outer) cluster radius, $R_{\rm h}$ is the half-mass radius, and $Q$ is the virial ratio.}
	\label{table2}
	\begin{tabular}{ccccccc}
		\hline 
        $\text{Model}$ & $R_{\rm t}$ & $R_{\rm c}$ & $R_{\rm h}$ & $Q$ \\
        \hline 
        Fractal & $\sim$ 3.0~pc & $\sim$ 4.0~pc & $\sim$ 3.0~pc & 0.5\\
        &  & $\sim$ 4.0~pc & $\sim$ 3.0~pc & 0.3\\
        &  & $\sim$ 3.0~pc & $\sim$ 2.0~pc & 0.5\\
        &  & $\sim$ 3.0~pc & $\sim$ 2.0~pc & 0.3\\
        &  & $\sim$ 2.0~pc & $\sim$ 1.0~pc & 0.5\\
        &  & $\sim$ 2.0~pc & $\sim$ 1.0~pc & 0.3\\
        &  & $\sim$ 1.0~pc & $\sim$ 0.7~pc & 0.5\\
        \hline
        Plummer & $\sim$ 3.0~pc & $\sim$ 3.0~pc & $\sim$ 2.0~pc & 0.5\\
        &  & $\sim$ 2.0~pc & $\sim$ 1.0~pc & 0.5\\
        &  & $\sim$ 1.0~pc & $\sim$ 0.7~pc & 0.5\\
        \hline
	\end{tabular}
\end{table}

\citet{Kim et al.2000} also consider the effective potential for a realistic Galactic tidal force. It includes the differential gravitational potential and the centrifugal potential in an acceleration form:
\begin{equation}
	\frac{\text{d}^{2} \bm{R_{\rm c}}}{\text{d}t^{2}}=\frac{2\text{G}M_{\rm c} \bm{R_{\rm c}}}{{R_{\rm G}}^{3}}  - \boldsymbol{\Omega}\ \times\ (\boldsymbol{\Omega}\ \times\ \bm{R_{\rm c}}),
\end{equation}
where $R_{\rm c}$ is the total cluster radius.  The first term is the differential gravitational potential and the second term is the centrifugal potential.

\subsection{Cluster Mass and IMF}

We simulate clusters of mass $\sim 2.0 \times 10^{4}$~M$_{\odot}$ from the best fit model for the Arches from \citet{Kim et al.2000}. We set the total number of stars to be $N=31000$, and randomly select stellar masses from the \citet{Maschberger2013} initial mass function between 0.01 and 100~M$_{\odot}$. This results in a total cluster mass $\sim 2.0 \times 10^{4}$~M$_{\odot}$.

Masses are initially distributed at random in the clusters (i.e. there is no primordial mass segregation, but we do examine if mass segregation occurs dynamically during the evolution).

\subsection{Initial distributions}

We use both Fractal \citep{Goodwin&Whitworth2004} and (spherical) Plummer \citep{Plummer1911,Aarseth et al.1974} distributions for the initial distributions.

Fractal initial conditions are chosen as a (hopefully) reasonable approximation to realistic substructured distributions that follow the turbulent gas in star forming Giant Molecular Clouds (GMCs) (although we note that in the environment of the GC it is not obvious that GMC structure would be the same as we observe in the outer Galaxy).  A Plummer sphere is chosen as it is a simple model that fits the relaxed distributions of older star clusters, as shown by e.g. \citet{Allison et al.2010} and \citet{Parker et al.2011} bound fractal distributions rapidly relax into a Plummer-like configuration in the absence of a strong tidal field (as we show below the same is true in strong tidal fields as long as the initial distribution is contained).

A Plummer sphere is a simple model similar to the current state of the Arches. It is defined by the total mass and a scale radius (we use the half-mass radius). Formally the Plummer sphere is infinite in extent, however if a truncation radius is set to be several scale radii then they are relatively stable. Plummer spheres are set-up using the prescription of \citet{Aarseth et al.1974}.

Fractals are constructed following \citet{Goodwin&Whitworth2004}. A box fractal is constructed in a cube and a sphere is cut from the cube and scaled to the desired total size. Velocity structure is produced by inheriting velocities (plus a small random component) from a parent during the generation of the box fractal. This produces locally correlated velocities which are then scaled to the desired total virial ratio. We use moderately substructured initial  distributions with fractal dimension $D=2.0$. 

The characteristic size/density of a Plummer sphere is set by the half-mass radius, $R_{\rm h}$, and the total cluster radius $R_{\rm c}$ is rather unimportant as long as it is several $R_{\rm h}$ (as density drops rapidly beyond the half-mass radius).  However, for fractals the important scale radius is the total cluster radius $R_{\rm c}$ as that contains all of the mass, the half-mass radius is poorly-defined and rather unhelpful as the mass distribution within $R_{\rm h}$ is clumpy.  Therefore while we quote both $R_{\rm h}$ and $R_{\rm c}$ for both Plummers and fractals the important radii are $R_{\rm h}$ for Plummers, and $R_{\rm c}$ for fractals.

{\bf A note on gas.}  Our simulations are purely $N$-body and do not include any contribution from gas left-over after star formation.  This is mainly a computational limitation in that even pure $N$-body calculations are expensive, and $N$-body plus hydro would be significantly more-so.  However, given the results of our simulations we return in the discussion to an argument that the star formation efficiency must have been high as any distribution must be contained within its tidal radius to survive.

\subsection{Initial internal energy}

We set the initial internal energy of our star clusters using the isolated global virial ratio, $Q$.  That is, the ratio of kinetic to potential energies of the clusters if they were in isolation.

For Plummer spheres we always use $Q=0.5$ (virialised), but for fractal distributions we use cool ($Q = 0.3$) and tepid\footnote{Note that $Q=0.5$ is a virial balance of energies, but due to the fractal distribution these regions are not in equilibrium.} ($Q=0.5$). Fractals in isolation will shrink in size by a factor of several, erasing substructure and reaching virial equilibrium \citep{Allison et al.2009a}.

\subsection{Summary of initial conditions}

We simulate $\sim 2.0 \times 10^{4}$~M$_{\odot}$ ($N=31000$) clusters for 2~Myr in a realistic strong tidal field 30~pc and 100~pc from the Galactic Centre. Our initial distributions are virialised Plummer spheres, and both cool and tepid $D=2.0$ fractal distributions. A detailed summary of the initial conditions that we use is given in tables~\ref{table1} and \ref{table2}.

\subsection{The mass segregation ratio, \texorpdfstring{$\Lambda_{\text{MSR}}$}{Lg}}
\label{section2.6}

Mass segregation is a more concentrated distribution of more massive stars than lower mass stars.  There are a number of ways of attempting to quantify if the massive stars are distributed differently, but probably the most useful is that of \citet{Allison et al.2009b} as it makes no assumptions about the underlying density distribution of stars or require a `centre' to be determined \citep{Parker&Goodwin2015}.

\citet{Allison et al.2009b} introduced the mass segregation ratio  $\Lambda_{\text{MSR}}$. The value of $\Lambda_{\text{MSR}}$ is a measure of how much more concentrated a particular set of the $N$ most massive stars compared to many sets of $N$ random stars of any mass.  The `length' of the distribution is the length of the minimum spanning tree (MST) between the $N$ members of a set. The value of $\Lambda_{\text{MSR}}$ is the ratio of the length of the MST of the $N$ most massive stars, $l_{\rm massive}$ to that of the average of many sets of $N$ random stars, $\langle l_{\rm norm} \big \rangle$
\begin{equation}
	\Lambda_{\rm MSR} = \frac{\big \langle l_{\rm norm} \big \rangle}{l_{\rm massive}} \pm \frac{\sigma_{\rm norm}}{l_{\rm massive}}.
\end{equation}
How unlikely it is that $l_{\rm massive}$ is drawn from the distribution of random values is given by the 1$\sigma$ standard deviation of $\langle l_{\rm norm} \big \rangle$.  If $\Lambda_{\text{MSR}} \sim 1$ (within the `errors') then there is no significant difference between the distributions of the most massive stars and random stars, when $\Lambda_{\text{MSR}}$ is significantly $>1$ then the most massive stars are more concentrated.

%%%%%%%%%%%%%%%%%%%%%%%%%%%%%%%%%%%%%%%%%%%%%%%%%%%%%%%%%%%%%%%%%%%%%%%

\section{Results}

We are interested in what initial conditions give rise to clusters that look similar to the Arches after $\sim 2$~Myr in a strong tidal field. First we will examine how clusters survive and evolve at 30~pc from the Galactic Centre (GC), and then compare this with similar initial conditions at 100~pc from the GC.

\subsection{Clusters at 30~pc from the GC}
\label{section3.1}

%%%%%%%%%%%%%%%%%%%%%%%%%%%%%%%%%%%%%%%%%%%%%%%%%%%%%%%%%%
\subsubsection{Well-contained clusters}

\begin{figure*}
\centering
    \includegraphics[angle=90,width=0.8\linewidth]{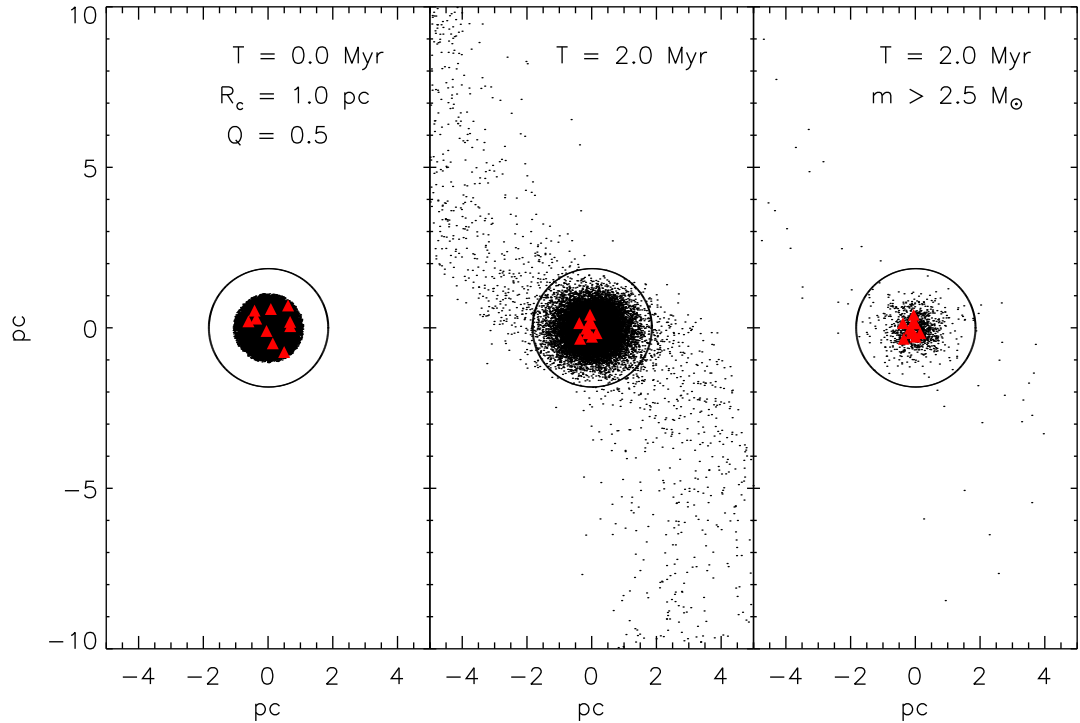}
	\caption{The evolution of an $R_{\rm c} = 1$~pc, virial ($Q=0.5$) Plummer cluster at 30~pc from the GC. The black dots are stars and the red triangles the 10 most massive stars in the cluster. The black circles are the nominal initial tidal radius ($R_{\rm t}$). The left and middle panels show all stars at 0 and 2~Myr, and the right figure shows only stars more massive than $\sim$ 2.5~M$_{\odot}$ at 2~Myr (an `observational' limit, see text for explanation).}
\label{fig01}
\end{figure*}

\begin{figure}
	\centering
    \includegraphics[angle=90,width=0.9\linewidth]{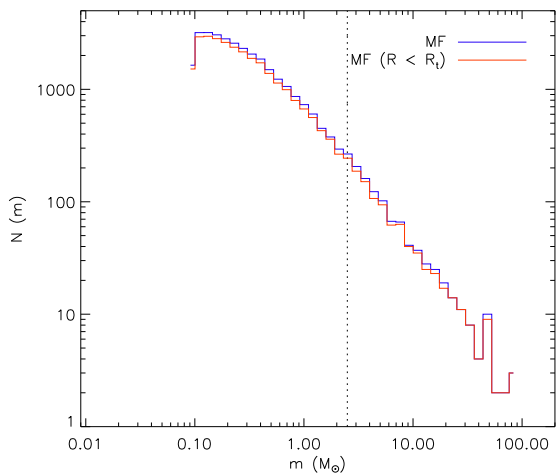}
	\caption{The mass functions (MFs) of a final distribution of Plummer star cluster (the right panel of Fig.\ref{fig01}). The blue solid line is all stars, and the red line is only stars inside the tidal radius ($R_{\rm t}$) of this cluster. The vertical dotted line shows the 2.5~M$_{\odot}$ `observational' selection limit.}
\label{fig02}
\end{figure}

\begin{figure*}
	\centering
    \includegraphics[angle=90,width=0.8\linewidth]{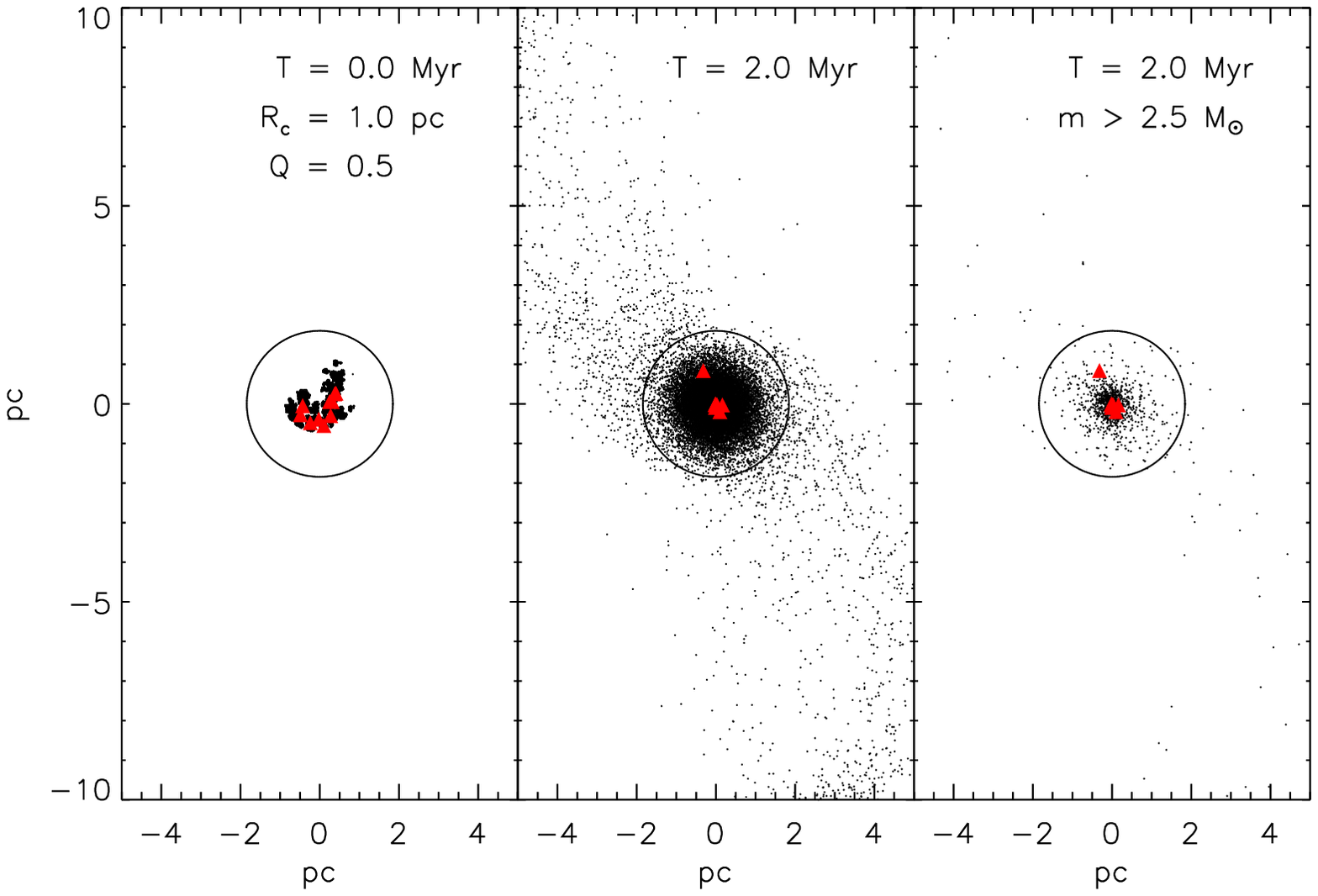}
	\caption{The evolution of an $R_{\rm c} = 1$~pc, tepid ($Q=0.5$) fractal cluster at 30~pc from the GC. The black dots are stars and the red triangles the 10 most massive stars in the cluster. The black circles are the nominal initial tidal radius ($R_{\rm t}$). The left and middle panels show all stars at 0 and 2~Myr, and the right figure shows only stars more massive than $\sim$ 2.5~M$_{\odot}$ at 2~Myr.}
\label{fig03}
\end{figure*}

\begin{figure}
	\centering
    \includegraphics[width=0.9\linewidth]{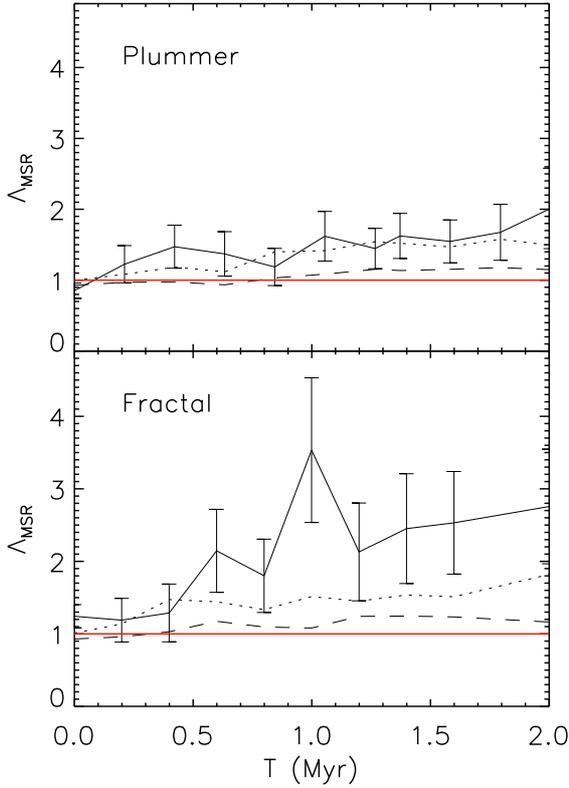}
	\caption{The evolution of $\Lambda_{\text{MSR}}$ as a function of time for the 10, 50, and 150 (solid line, dotted, and dashed) most massive stars in Fig.~\ref{fig01} and \ref{fig03}. The upper panel shows a Plummer star cluster, and the lower panel shows a fractal star cluster. The error bar means 1$\sigma$ error. We plot error bars only for 10 most massive stars for clarity. The red solid line shows no mass segregation ($\Lambda_{\text{MSR}}$ = 1).}
\label{fig04}
\end{figure}

\begin{figure*}
	\centering
	\includegraphics[width=0.9\linewidth]{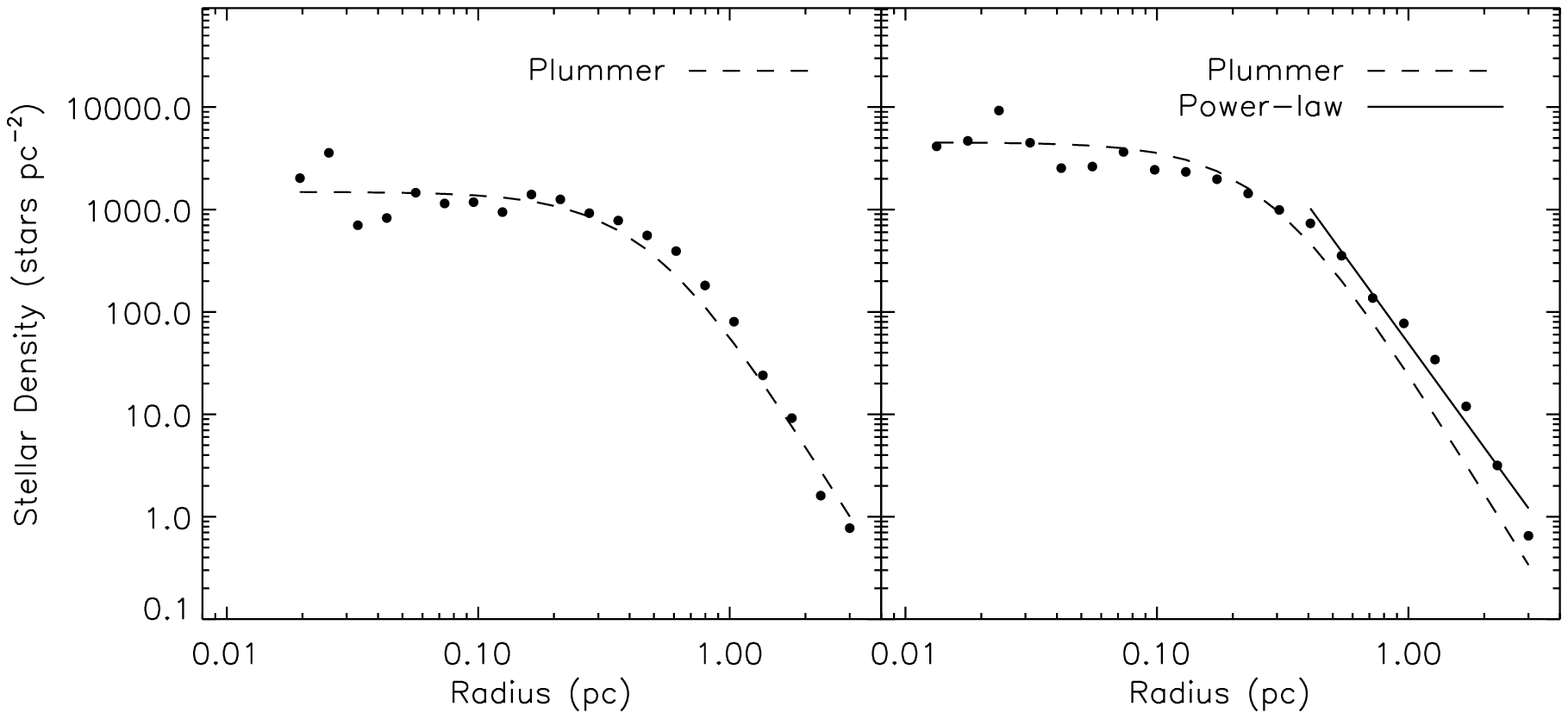}
	\caption{Final density distributions inside the tidal radius of stars $> 2.5 M_\odot$ shown by the dots. The initial Plummer sphere the is on the left (Fig.~\ref{fig01}), and initial fractal (Fig.~\ref{fig03}) is on the right.  The dashed lines are Plummer models, and the black line is a power-law slope.}
\label{fig05}
\end{figure*}

We define a `well-contained' cluster to be one which is initially well within its initial tidal radius (i.e. a cluster that would probably be expected to survive).  

{\bf A dense, virialised Plummer sphere.} 
Let us first examine the evolution of `standard' initial conditions (i.e. initial conditions that are close to the currently observed state of the Arches). That is: a virialised Plummer sphere whose initial size is smaller than its (nominal) tidal radius ($R_{\rm c} = 1$~pc, and $R_{\rm h} \sim$ 0.7~pc ).

In Fig.~\ref{fig01}, we show the initial conditions at $T=0$~Myr in the left panel, and the `final' state at $T=2$~Myr in the middle panel (we will explain the right panel shortly). The majority of stars are shown by black dots, and the 10 most massive stars by red triangles.

The black circle around the cluster shows the size of the {\em initial} tidal radius ($R_{\rm t}$). As stars escape beyond the tidal radius the interior mass decreases and the tidal radius decreases, but for ease of comparison between panels we keep the circle at the size of the initial tidal radius.

The view point in all of our figures similar to Fig.~\ref{fig01} is viewing down from above the Galactic plane (i.e. viewing the plane of the orbit).  It is worth noting that this is not the view we have of the Arches cluster and the GC as we lie in the same plane, rather it is the view one might have of nuclear clusters in face-on galaxies.  Moving to a view-point similar to our own adds many complications as there are many viewing angles we might have.  In this paper we only view from above the Galactic plane (but we will return to the effect of viewing angles in a later paper).

In 2~Myr the dense Plummer sphere has not evolved significantly. In the middle panel it is obvious by visual inspection that two tidal tails have formed, and also that the massive stars have dynamically mass segregated. But there is no very significant difference between the cluster at 0~Myr and 2~Myr. The fraction of stars lost over the tidal boundary is roughly 7 per cent of the stars by number (but somewhat less by mass, see below).

Interestingly, \citet{Hosek et al.2015} find that the Arches cluster does not have an observable tidal tail. At face value this might rule out the Arches being at 30~pc from the GC as our simulated cluster quite clearly has a tidal tail. However, \citet{Hosek et al.2015} are only able to observe stars more massive than $\sim$ 2.5~M$_{\odot}$. When we apply this observational limit in the right panel of Fig.~\ref{fig01} we see that the tidal tail is now barely visible ($\sim 96$ per cent of stars in the tidal tail are $<$2.5~M$_{\odot}$).

This is mostly due to the vast majority of stars in our IMF being $<$2.5~M$_{\odot}$, but is enhanced by dynamical mass segregation (i.e. none of the most massive stars are in the tidal tail).

Mass segregation causes a difference between the half-mass radius measured in stars $>$2.5~M$_{\odot}$ of $\sim 0.5$~pc, compared to the `true' half-mass radius from all stars of $\sim 0.6$~pc. This suggests that a half-mass radius measured from only intermediate and massive stars could well under-estimate the true half-mass radius by $\sim 20$ per cent.

Fig. \ref{fig02} shows the mass functions of all stars (blue histogram), and only stars within the nominal tidal radius (red histogram). The mass function of stars within the tidal radius is very slightly different with slightly fewer low-mass stars than in the full IMF. However this would be essentially impossible to actually detect as (a) the difference is very small, (b) the difference occurs at the very low-mass/low-luminosity end of the mass function that would be extremely difficult to observe, and (c) once away from the `cluster' contamination from `background' stars would be almost impossible to disentangle.

\bigskip

{\bf A dense, tepid fractal cluster.} Next, we run a simulation with clumpy substructure, with a fractal dimension $D=2.0$ and virial ratio $Q=0.5$. The total radius of this fractal distribution is $R_{\rm c} = 1$~pc, i.e. the same total radius as the dense Plummer sphere above. 

In Fig.~\ref{fig03} we show the evolution of this dense and tepid fractal star cluster to compare directly with Fig.~\ref{fig01}. In the left panel we can see a clumpy and substructured star cluster. The middle panel shows the total final distribution at $\sim 2$~Myr, and right panel shows the final distribution with the `observational' 2.5~M$_\odot$ cut. In this case, $\sim$ 10 per cent of stars (by number) have escaped from the star cluster, and $\sim$ 96 per cent of the stars in the tidal tails are < 2.5~M$_{\odot}$.

It is worth noting that the initial fractal distribution does not seem quite centred in the tidal radius circle (this becomes more obvious in some later figures). This is an artifact of the fractal generation procedure as it is a sphere cut from a cube whose centre of mass is often not at the centre of the sphere. The centre of mass of the fractal {\em is} at the centre of the tidal radius circle, although this is sometimes not obvious to the eye.

The fractal distribution rapidly evolves into a compact, smooth and spherical star cluster within the tidal radius (cf. \citealt{Allison et al.2009a}). Again, there is a clear tidal tail if all stars are observed, but one that is barely present when applying the $\sim$ 2.5~M$_{\odot}$ cut.  And again, the half-mass radius measure for all stars or stars $>$ 2.5~M$_{\odot}$ is different (this time 0.5~pc for all stars, and 0.4~pc for the more massive stars).

\begin{figure*}
	\centering
	\includegraphics[angle=90,width=0.8\linewidth]{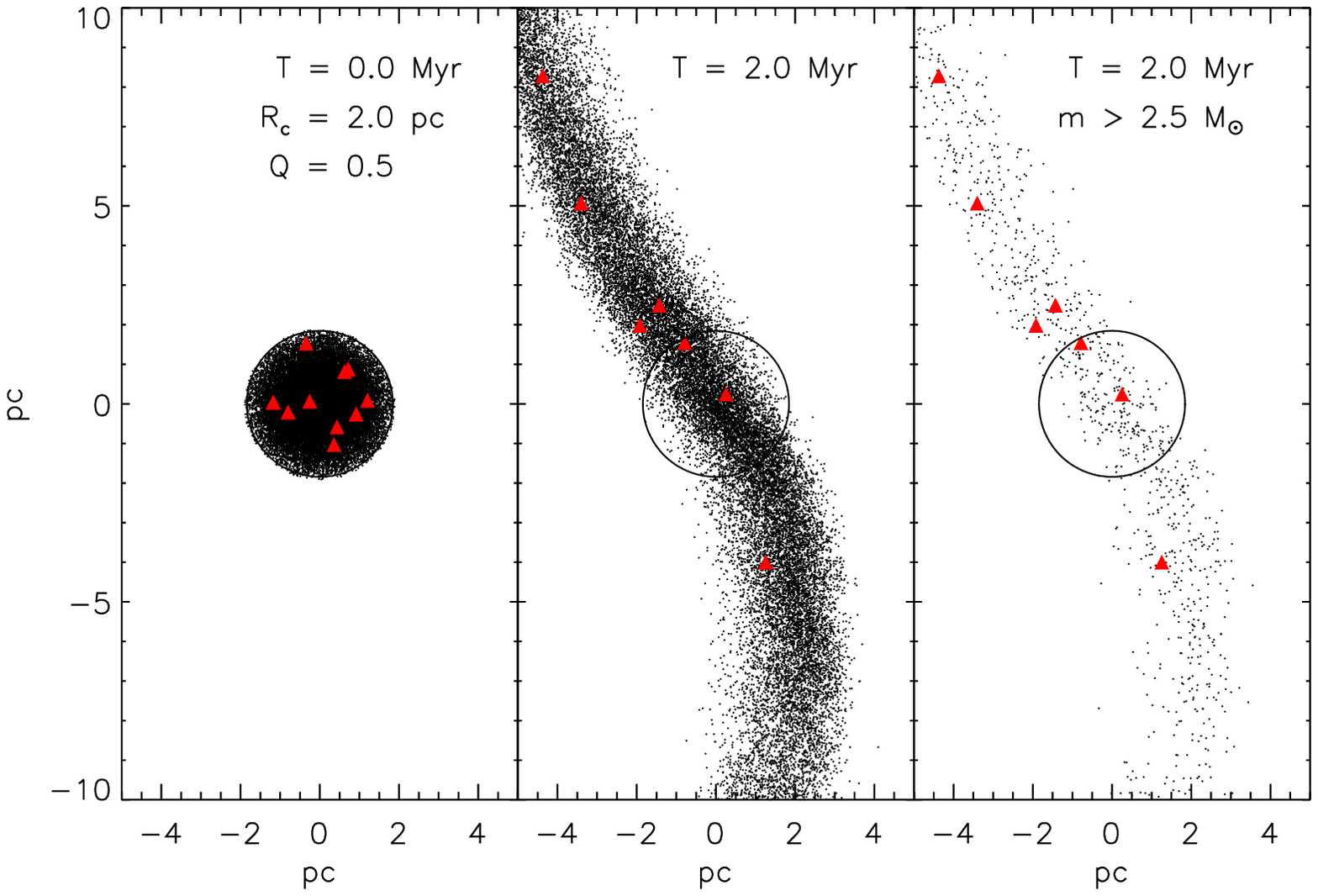}
	\caption{The evolution of an $R_{\rm c} = 2$~pc, virial ($Q=0.5$) Plummer cluster at 30~pc from the GC. The  black dots are stars and the red triangles the 10 most massive stars in the cluster. The black circles are the nominal initial tidal radius ($R_{t}$). The left and middle panels show all stars at 0 and 2~Myr, and the right figure shows only stars more massive than $\sim$ 2.5~M$_{\odot}$ at 2~Myr.} 
\label{fig06}
\end{figure*}

Note that the fractal distribution decreases in size.  A fractal distribution, even with a `virial' energy balance is not in equilibrium.  It will undergo violent relaxation which erases substructure and redistributes the potential energy in the substructure.  This is explained in detail in \citet{Allison et al.2009a}, but essentially the potential energy of a distribution is $\eta GM_{\rm c}^2/R$ where $R$ is a scale radius, and $\eta$ is a constant that depends on the density distribution: for a Plummer sphere $\eta \sim 0.75$ if $R$ is the Plummer radius, but in a very clumpy fractal $\eta \sim 1.5$.  Therefore if the virial ratio remains the same $R$ must fall by a factor of two after violent relaxation from a fractal to a Plummer-like distribution as $\eta$ has changed.  (This is rather stochastic as the exact value of $\eta$ depends on the particular details of each fractal realisation.)

The final state at 2~Myr of the dense Plummer sphere and the dense fractal are qualitatively and quantitatively very similar. After erasing the initial substructure, it is essentially impossible to distinguish the simulations at 2~Myr. (The particular fractal initial conditions we use give rise to a slightly denser final cluster, but we could easily `fine tune' either set of initial conditions to produce an almost identical final cluster.)

\bigskip

{\bf Mass segregation in high-density clusters.}  Fig.~\ref{fig04} shows the evolution of mass segregation as measured by $\Lambda_{\text{MSR}}$ (see Sec.~\ref{section2.6}) as a function of time for four subsets of $N_{\text{MST}}$ = 10, 50, and 150 (solid line, dotted, and dashed) most massive stars in the Plummer (upper panel) and fractal (lower panel) clusters\footnote{The masses of 10th, 50th, and 150th most massive star are $\sim$ 53~M$_{\odot}$, $\sim$ 25~M$_{\odot}$, and $\sim$ 11~M$_{\odot}$, respectively.}. To determine $\Lambda_{\text{MSR}}$ we use stars more massive than 2.5~M$_{\odot}$ that are within two tidal radii.

In both cases both the 10 and 50 most massive stars dynamically mass segregate in 2~Myr (full line and dotted line respectively). The fractal shows {\em slightly} more mass segregation than the Plummer sphere, but the difference is not particularly significant.

It is worth noting that in Fig.~\ref{fig01} we show only one realisation of the initial conditions.  Because Plummer spheres are simple, spherical models differences between different realisations are small and are only different in the initial locations of the most massive stars. Fractals, however, have stochasticity in the initial (position and velocity) structure as well as the placement of the most massive stars, therefore they show much more variation between different realisations (see e.g. \citealt{Allison et al.2010}).  At these densities, whilst there are differences between different realisations, the high density of the fractal initial conditions means that later dynamical evolution dominates over stochasticity in the initial conditions and all realisations of even the fractals are fairly similar at 2~Myr and what we have plotted is a  typical example of the evolution of $\Lambda_{\text{MSR}}$.

\subsubsection{Density profiles}

\begin{figure*}
	\centering
	\includegraphics[angle=90,width=0.8\linewidth]{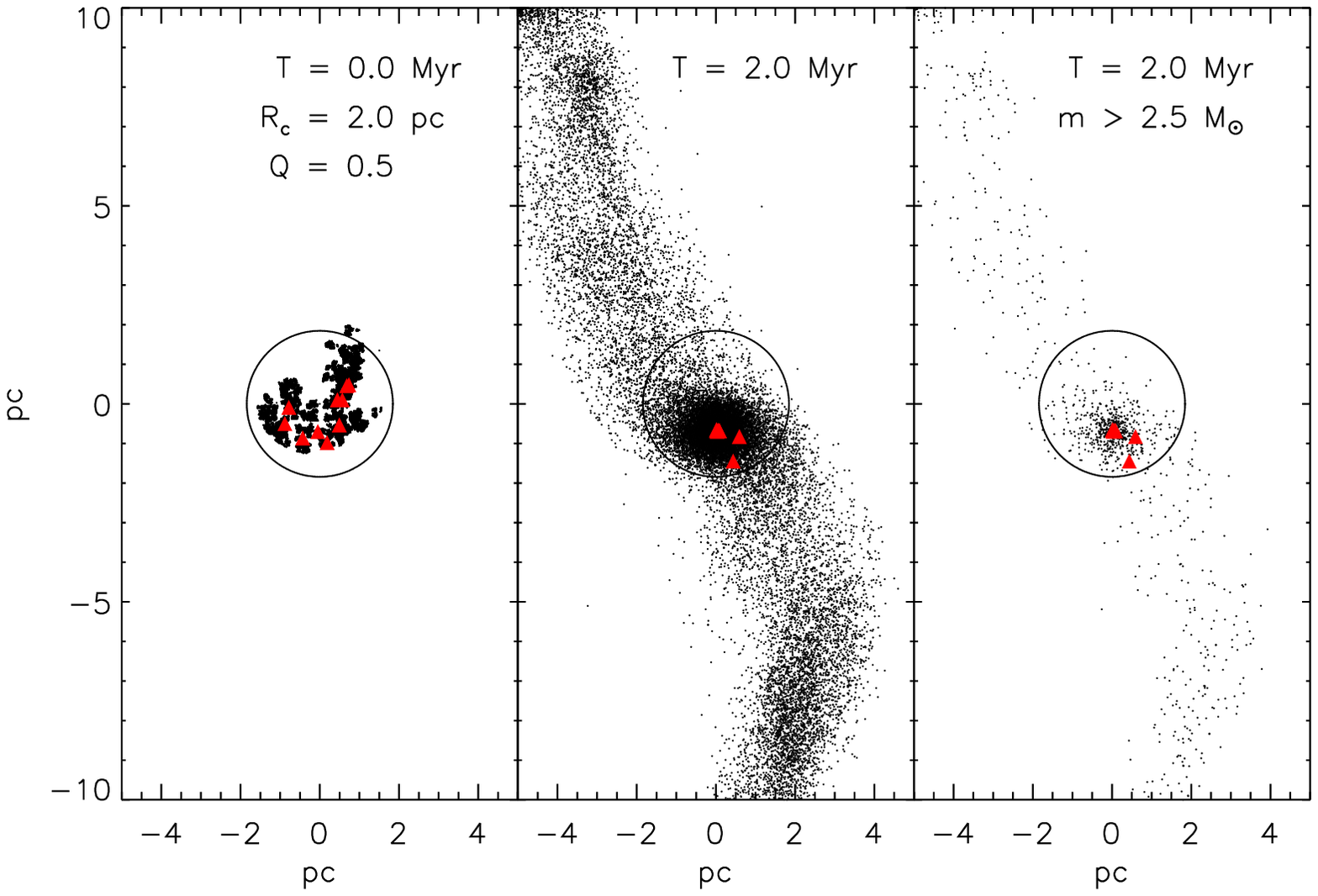}
	\caption{The evolution of an $R_{\rm c} = 2$~pc, tepid ($Q=0.5$) fractal cluster at 30~pc from the GC. The black dots are stars and the red triangles the 10 most massive stars in the cluster. The black circles are the nominal initial tidal radius ($R_{t}$). The left and middle panels show all stars at 0 and 2~Myr, and the right figure shows only stars more massive than $\sim$ 2.5~M$_{\odot}$ at 2~Myr.}
\label{fig07}
\end{figure*}

Another slight, but potentially observable, difference between the initial Plummer sphere and initial fractal are the density profiles of the final clusters.

In Fig.~\ref{fig05} we show the final density profiles of the initially well-contained Plummer sphere on the left, and the initially well-contained fractal on the right.  To match what might be observable we determine the profiles only from stars $> 2.5$~M$_{\odot}$, and inside the tidal radius ($R_{\rm t} \sim$ 2~pc). 

On the left, is the profile of the cluster resulting from the initial Plummer sphere which has kept a density distribution very similar to that with which it started (with some, unsurprising, evidence of tidal truncation at large radii).  

On the right, is the profile of the initially fractal region (which underwent violent relaxation and erased its substructure) and its final density profile is somewhat steeper with a power-law decline, and a Plummer model is not a particularly good fit to the profile.

\citet{Hosek et al.2015} find that the outer density profile of the Arches is fitted well by a power-law (see their Fig.~14) rather than a King model (i.e. a tidally truncated Plummer-like model).  Their fit is more similar to our results for what was initially a fractal, but we are hesitant to make too much of this as it is quite possible that small adjustments to the initial Plummer sphere's density profile could reproduce the observations as well.  However, it is interesting that the observed profile is what is expected of a post-violent relaxation clumpy distribution.

\subsubsection{Just-contained clusters}

\begin{figure*}
	\centering
        \includegraphics[angle=90,width=0.8\linewidth]{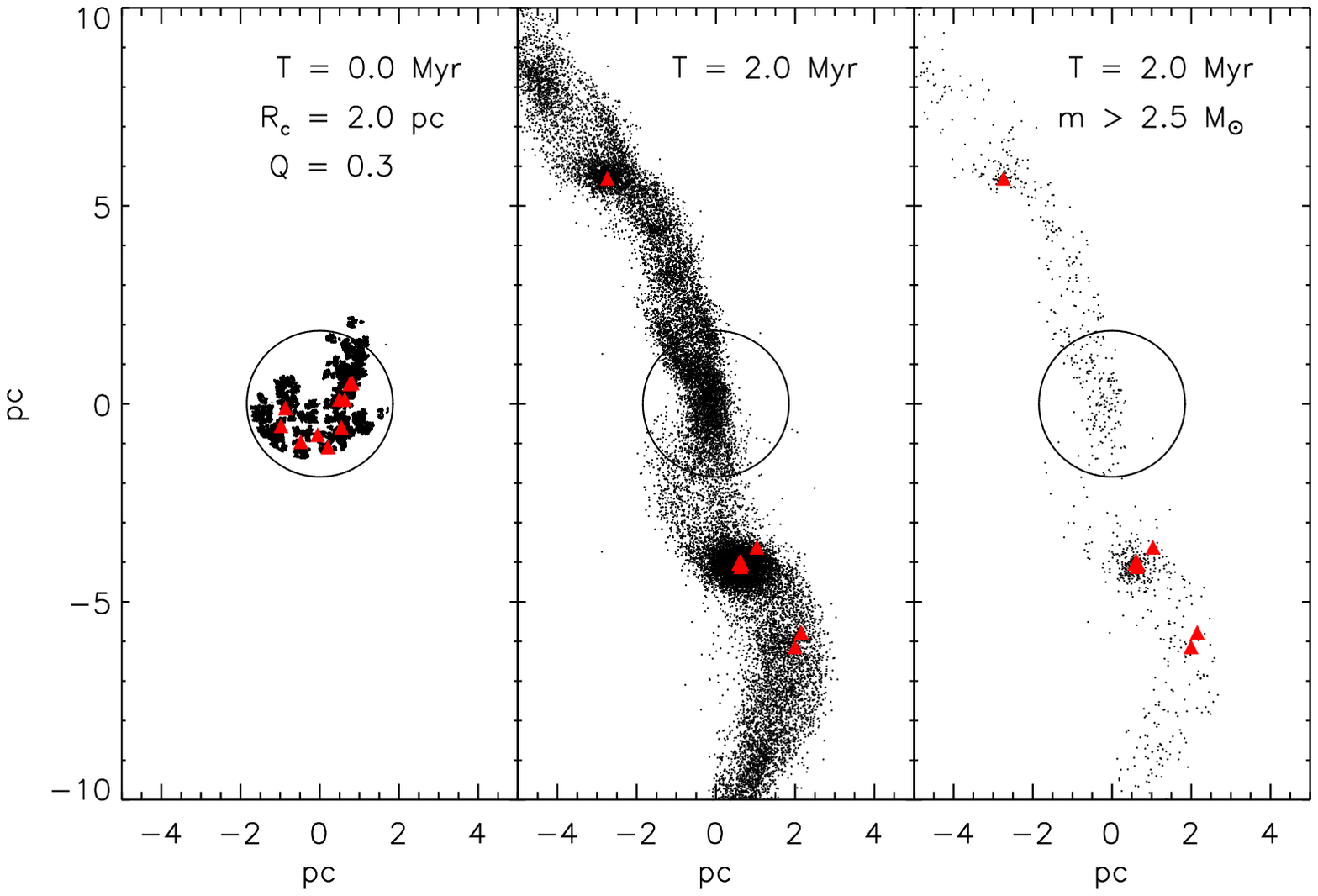}
	\caption{The evolution of an $R_{\rm c} = 2$~pc, cool ($Q=0.3$) fractal cluster at 30~pc from the GC. The black dots are stars and the red triangles the 10 most massive stars in the cluster. The black circles are the nominal initial tidal radius ($R_{t}$). The left and middle panels show all stars at 0 and 2~Myr, and the right figure shows only stars more massive than $\sim$ 2.5~M$_{\odot}$ at 2~Myr.}
\label{fig08}
\end{figure*}

\begin{figure}
	\centering \includegraphics[width=0.9\linewidth]{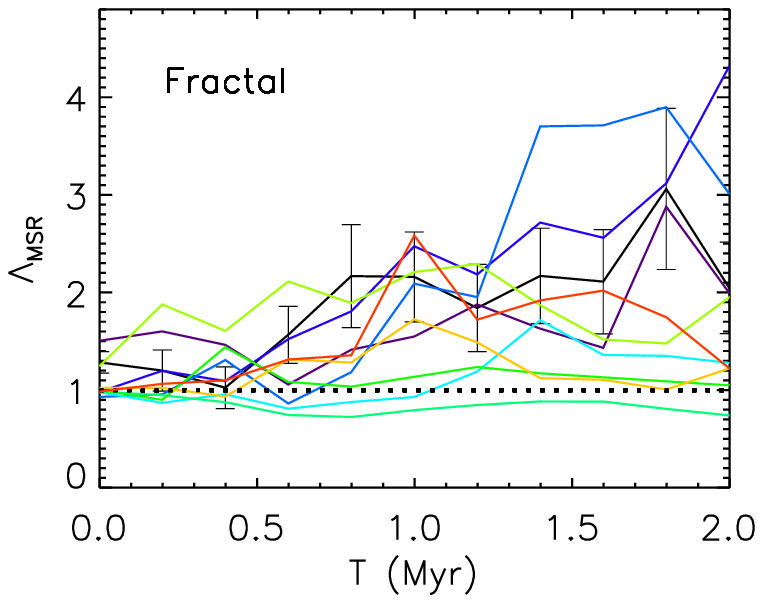}
	\caption{The evolution of $\Lambda_{\text{MSR}}$ as a function of time for 10 most massive stars with 10 ensembles of tepid ($Q = 0.5$) fractal initial condition at 30~pc from the GC when $R \sim R_{t}$. Coloured solid lines mean each ensemble. Black dotted line indicates no mass segregation. We plot the 1~$\sigma$ error bars only one case for clarity.}
\label{fig09}
\end{figure} 

It is no particular surprise that bound clusters initially well-contained within their tidal radius are able to survive. Therefore, we now examine clusters that are {\em just} contained within their tidal radius (and so are larger and less dense than the `well-contained' clusters considered above).

{\bf A virialised just-contained Plummer sphere.}
In Fig.~\ref{fig06} we show the initial and final states of an intermediate density virialised Plummer sphere with total radius $R_{\rm c} = 2$~pc (cf. Fig.~\ref{fig01}). It is clear from Fig.~\ref{fig06} that the initial distribution fills the initial tidal radius (although the circle indicating the tidal radius in Fig.~\ref{fig06} is rather hard to see).

After 2~Myr, the just-contained virialised Plummer sphere has not survived as a cluster (middle and right panels of Fig.~\ref{fig06}). In this case, as the intermediate-density virialised Plummer sphere initially fills the tidal radius,  stars in the outskirts are more affected by Galactic tidal field than those in the inner region, so stars in the outskirts can escape through the Lagrange points. The more stars that escape, the smaller the tidal radius becomes, and the mass loss becomes more rapid, and so-on.

\begin{figure*}
	\centering
    \includegraphics[angle=90,width=0.8\linewidth]{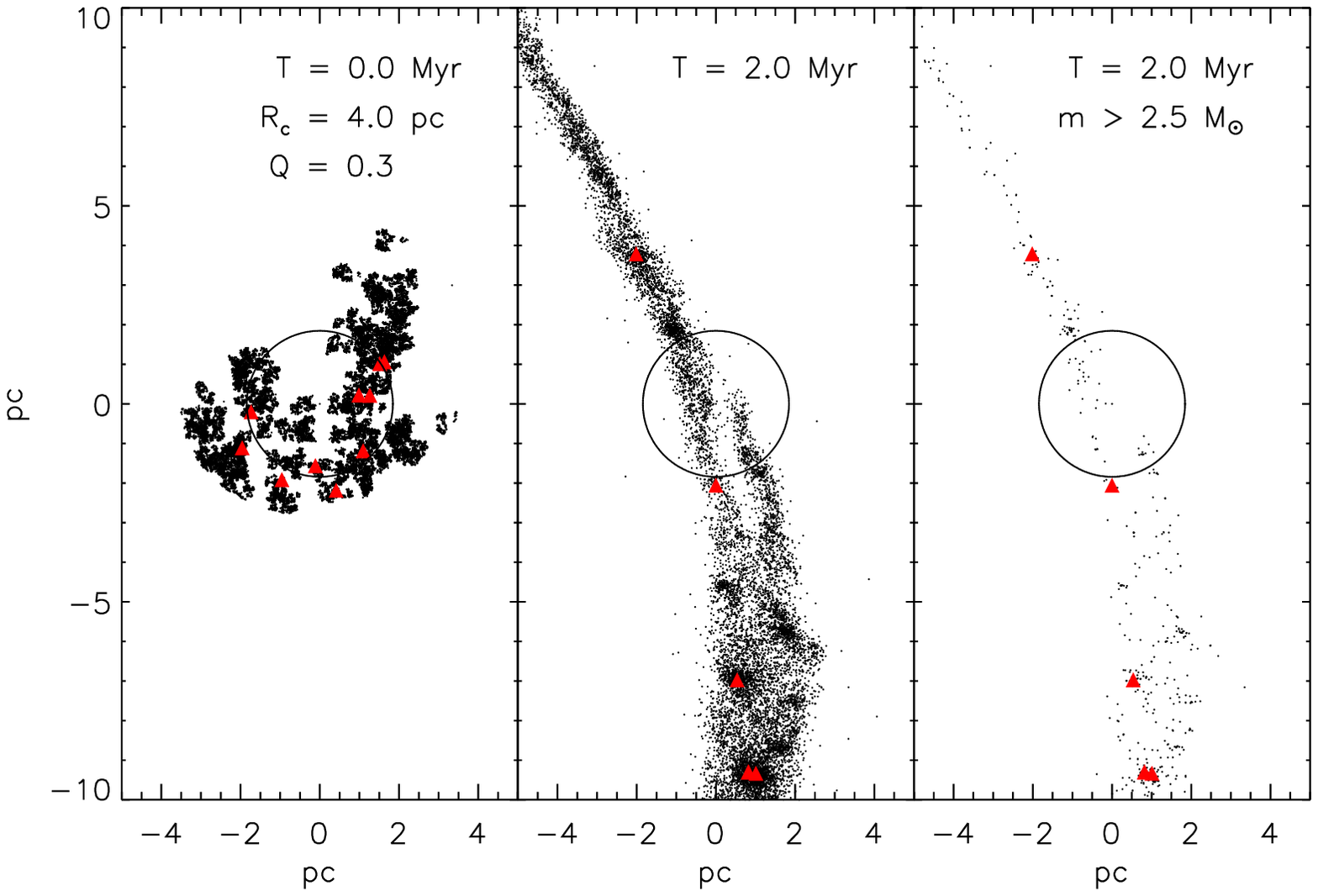}
	\caption{The evolution of an $R_{\rm c} = 4$~pc, cool ($Q=0.3$) fractal cluster at 30~pc from the GC. The black dots are stars and the red triangles the 10 most massive stars in the cluster. The black circles are the nominal initial tidal radius ($R_{t}$). The left and middle panels show all stars at 0 and 2~Myr, and the right figure shows only stars more massive than $\sim$ 2.5~M$_{\odot}$ at 2~Myr.}
\label{fig10}
\end{figure*}

\begin{figure*}
	\centering
	\includegraphics[angle=90,width=0.8\linewidth]{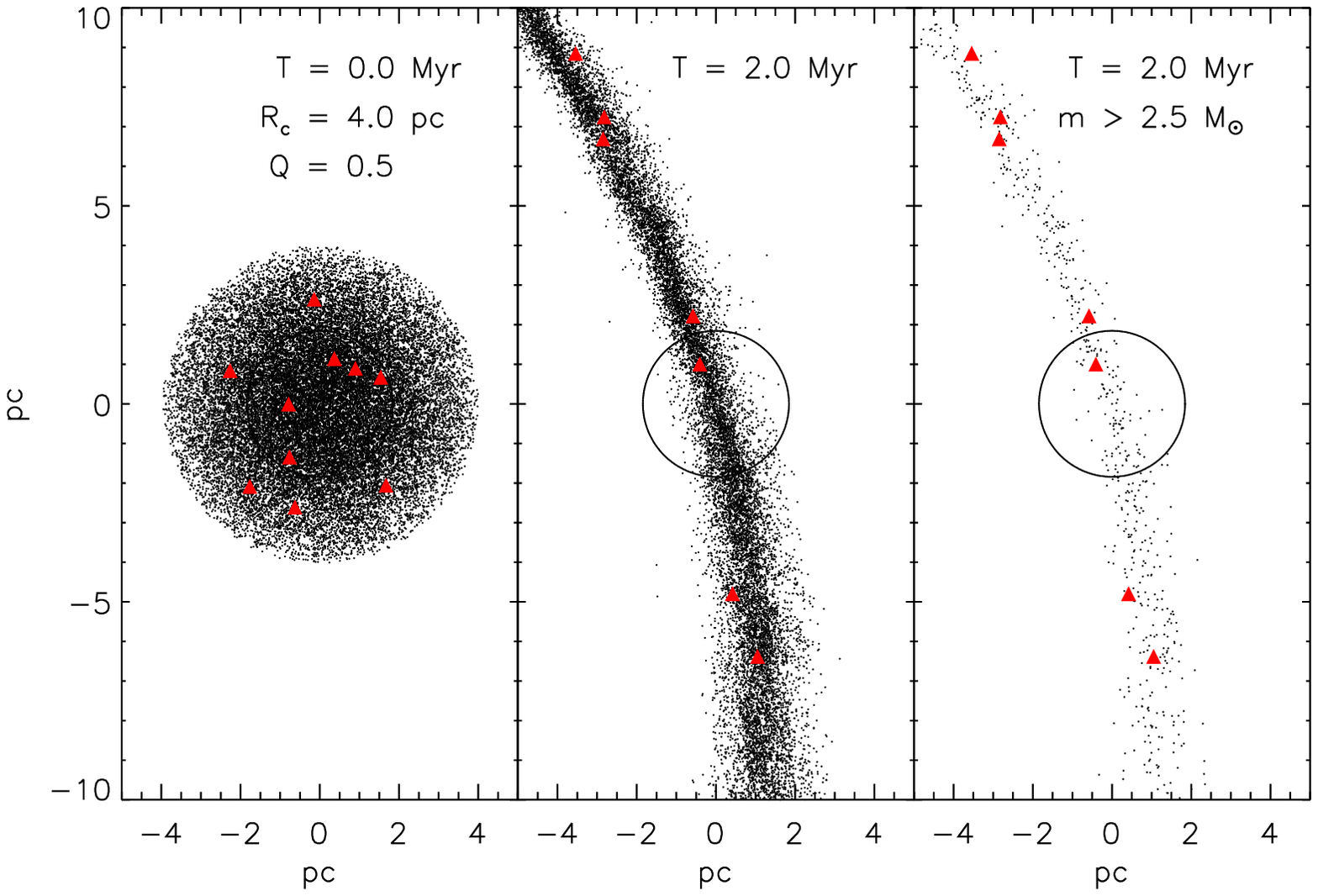}
	\caption{The evolution of an $R_{\rm c} = 4$~pc, virial ($Q=0.5$) Plummer cluster at 30~pc from the GC. The black dots are stars and the red triangles the 10 most massive stars in the cluster. The black circles are the nominal initial tidal radius ($R_{\rm t}$). The left and middle panels show all stars at 0 and 2~Myr, and the right figure shows only stars more massive than $\sim$ 2.5~M$_{\odot}$ at 2~Myr.}
\label{fig11}
\end{figure*}

{\bf A tepid just-contained fractal.} 
In Fig.~\ref{fig07}  we show the the initial and final states of an intermediate density tepid ($Q=0.5$) fractal with total radius 2~pc (compare with Fig.~\ref{fig03}), we can see a surviving star cluster within the tidal radius unlike Fig.~\ref{fig06}. A single cluster remains after 2~Myr which is not as mass segregated as in the high-density case (compare with Fig.~\ref{fig03}). In this case, $\sim$ 53 per cent of stars (by number) escape from the star cluster, and again $\sim$ 96 per cent of the stars in the tidal tails are < 2.5~M$_{\odot}$.

The reason the intermediate-density fractal could survive is that it undergoes a rapid collapse and erasure of substructure resulting in a denser final Plummer-like sphere than the initial just-contained Plummer sphere which was destroyed.

Here we have an apparent difference between Plummer sphere and fractal initial conditions, but it is (again) quite minor.  A slightly more extended fractal would be destroyed (see below), and a slightly more compact Plummer sphere would survive.  But, because of the collapse of fractals they are able to survive with slightly more extended initial distributions.

{\bf Mass segregation.}
As noted above, for the initial Plummer spheres, just-contained initial conditions cannot survive for 2~Myr. In the fractal case where they can survive, one difference between initially well-contained and just-contained clusters is the degree of mass segregation observed after 2~Myr (a visual inspection of Fig.~\ref{fig07} shows much less apparent mass segregation than in Fig.~\ref{fig03}).

Earlier we discussed only one realisation of a fractal cluster as the dynamical age of the initially well-contained clusters means there is little variation between realisations.  However, there is rather more variation in the final states of the lower-density and just-contained clusters as they are dynamically younger at 2~Myr than well-contained clusters.

Therefore, in Fig.~\ref{fig09} we plot the evolution of $\Lambda_{\text{MSR}}$ for each of 10 different realisations fractal initial conditions: each differently coloured line is a different realisation (in each case $\Lambda_{\text{MSR}}$ is determined for the 10 most massive stars compared to stars $2.5$~M$_\odot$ within two tidal radii, as above).

In Fig.~\ref{fig09} we see that there is a lot of variation in $\Lambda_{\text{MSR}}$ between realisations after $\sim 1$~Myr with around half of clusters showing very significant mass segregation signatures, and half no signature.  This is due to both the ejection of a high-mass star, or the constant formation and destruction of higher-order Trapezium-like multiples (see \citealt{Allison&Goodwin2011}).

\bigskip

{\bf A cool just-contained fractal.}
In Fig.~\ref{fig08} we show the the initial and final states of the same fractal distribution as in Fig.~\ref{fig07}, but having reduced the virial energy ratio to $Q=0.3$.

As this fractal is cool, in the absence of a strong tidal field, it would collapse to a denser state than the tepid fractal we discussed above. However, in the middle panel of Fig.~\ref{fig08} we see the rather unanticipated result that rather than forming a single cluster with tidal `arms', the stars are spread along the orbit, but with significant over-densities.  

In particular, just below the `starting point' (at roughly $-5$~pc in the middle panel) is a significant cluster containing 6 of the most massive stars. Its total mass is $\sim$ 5500~M${_\odot}$, with over half of its mass ($\sim$ 2800~M$_{\odot}$) in stars $> 2.5$~M$_{\odot}$.  It is this cluster that is the `remnant' of the main cluster, having lost a significant amount of its mass now spread along the orbit. There is an another small remnant at roughly +5~pc in the middle panel, its total mass is $\sim$ 1300~M$_{\odot}$ with again roughly half its mass ($\sim$ 700~M$_{\odot}$) in stars $> 2.5$~M$_{\odot}$.

It is worth noting that if the two surviving sub-clumps were observed and thought to be discrete `units' of star formation they would appear to have top-heavy IMFs (our initial global IMFs have $\sim 42$ per cent by mass in stars $> 2.5$~M$_\odot$).  In the context of these simulations we know that they are mass segregated and tidally stripped subunits from a `normal' IMF, but this may well not be apparent when observing a single (late) point in the evolution of the region.

The reason for this very different behaviour is that a collapsing cluster will `bounce', i.e. in the destruction of substructure and the attempt to reach virial equilibrium causes a deep collapse after which the cluster `bounces'. As mentioned above, fractal clusters will undergo violent relaxation and erase their substructure causing them to collapse.  However, they do not immediately relax into a smooth, virialised distribution, rather they collapse down to an over-dense state, and then re-expand and `bounce' for a while with the virial ratio oscillating around $Q=0.5$ (see e.g. \citealt{Allison et al.2009a,Smith et al.2011}). Depending on the exact time of an observation the virial ratio varies from about 0.3 to 0.5 to 0.7, back to 0.5 etc. This oscillation is enhanced by rapid dynamical mass segregation that gives energy to low-mass stars \citep{Allison et al.2009a}.  With no tidal field this oscillation dies away and the cluster stabilises \citep[see e.g. Fig. 10 in]{Smith et al.2011}, but in a strong tidal field the re-expansion takes some stars outside the tidal radius where they can be stripped.

If we were to observe the state at 2~Myr it would be extremely difficult (essentially impossible) to reconstruct the initial conditions. Probably the most obvious conclusion one would draw from observing the structure in the middle panel of Fig.~\ref{fig08} is that the initial star formation event was extended over $>10$~pc, and it would probably not be obvious that all of these stars had formed within a 2~pc radius.

This illustrates that it can be very difficult to `guess' the results of evolution of a system in a strong tidal field.  Our assumption before running this simulation was that a cool fractal within the nominal tidal radius would survive, and would probably produce a cluster that looked similar to the well-contained initial conditions.

\subsubsection{An overflowing, cool fractal cluster}  

Following well-contained, and just-contained initial conditions we now consider initial conditions that `overflow' (i.e. are larger than) the nominal initial tidal radius.

As described above, in no (or weak) tidal fields cool (subvirial) fractals can collapse by factors of several in radius forming dense and mass segregated clusters very quickly (on timescales of $\sim$ 1~Myr).  Therefore, it is interesting to investigate if a cool structure which is initially overflowing the tidal radius can collapse to within the tidal radius before the tidal field destroys it.  We take the same cool ($Q=0.3$) fractal distribution as used above and increase the radius (by a factor of 2) to 4~pc. 

In Fig.~\ref{fig10} we show the initial conditions of this low-density fractal (left panel), and the state after 2~Myr (middle and right panels with the right panel again showing only the stars
>2.5~M$_\odot$). 

What is clear from Fig.~\ref{fig10} is that the substructured initial cluster is completely destroyed by 2~Myr. The main reason for this is that our nominal circular tidal radius drawn on all figures assumes a concentration of mass within that radius. When starting with a clumpy mass distribution larger than the nominal tidal radius the effect is to divide the cluster into subregions each with their own tidal radius: i.e. if the density within a group is large enough then that group may survive, but the entire distribution is not a single entity, essentially there is a very significant shear over an 8~pc region at 30~pc from the GC (i.e. between 26 and 34 pc from the GC). The fractal cannot survive as a single entity to collapse, but some sub-regions are able to survive (e.g. a significant over-density at around $-10$~pc in the middle panel containing two massive stars, and a smaller over-density at around $-7$~pc containing one).

\begin{figure*}
	\centering
	\includegraphics[angle=90,width=0.8\linewidth]{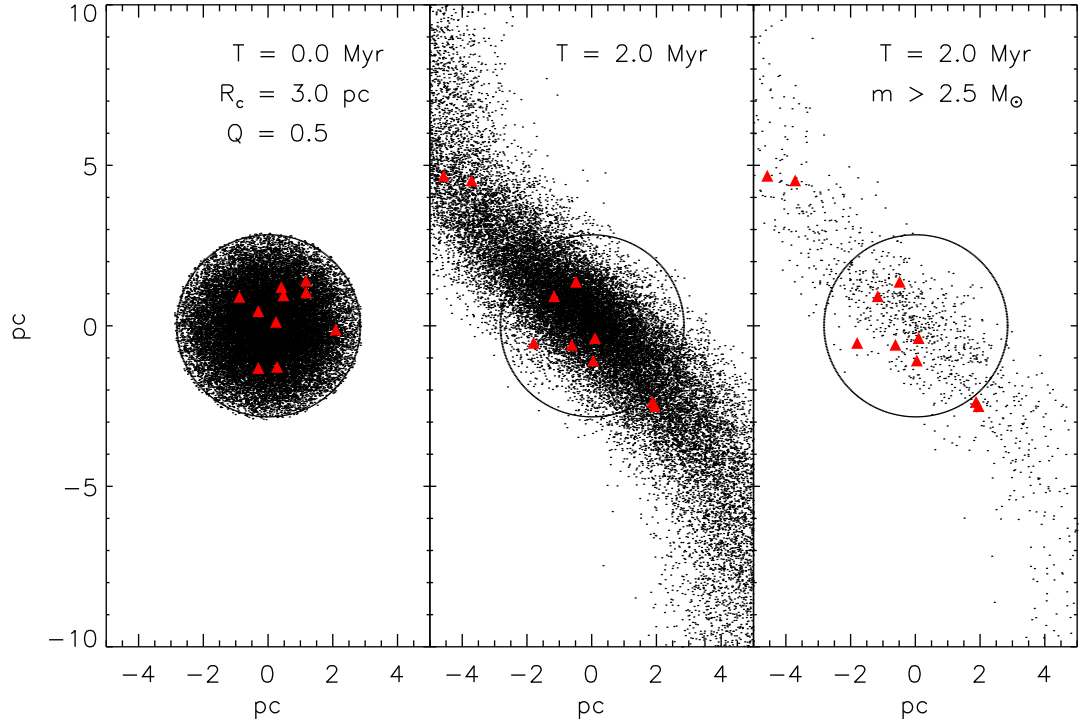}
	\caption{The evolution of an $R_{\rm c} = 3$~pc, virial ($Q=0.5$) Plummer cluster at 100~pc from the GC. The black dots are stars and the red triangles the 10 most massive stars in the cluster. The black circles are the nominal initial tidal radius ($R_{\rm t}$). The left and middle panels show all stars at 0 and 2~Myr, and the right figure shows only stars more massive than $\sim$ 2.5~M$_{\odot}$ at 2~Myr.}
\label{fig12}
\end{figure*}

\begin{figure*}
	\centering
	\includegraphics[angle=90,width=0.8\linewidth]{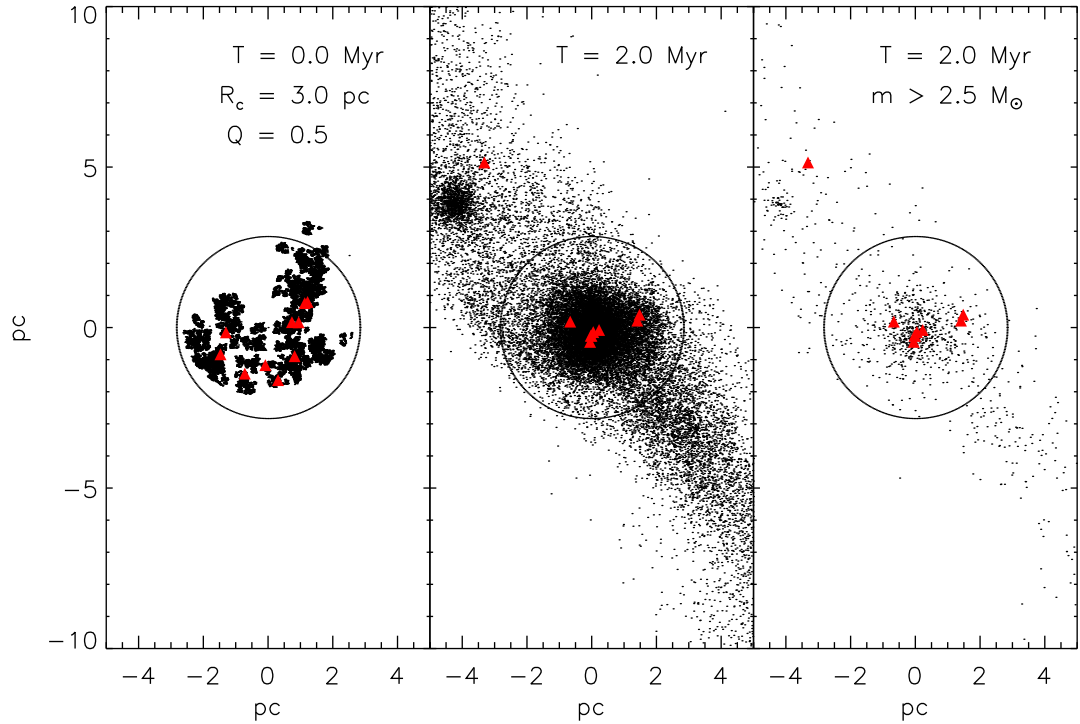}
	\caption{The evolution of an $R_{\rm c} = 3$~pc, tepid ($Q=0.5$) fractal cluster at 100~pc from the GC. The black dots are stars and the red triangles the 10 most massive stars in the cluster. The black circles are the nominal initial tidal radius ($R_{\rm t}$). The left and middle panels show all stars at 0 and 2~Myr, and the right figure shows only stars more massive than $\sim$ 2.5~M$_{\odot}$ at 2~Myr.}
\label{fig13}
\end{figure*}

Similarly, we simulate a virialised Plummer sphere which is larger than its tidal radius as illustrated in Fig.~\ref{fig11}. This is also very rapidly destroyed by the tidal field for exactly the same reason. However, as the initial Plummer distribution was much smoother than the fractal no structures or over-densities remain and the final state is smoothly distributed around the orbit. This shows that structure/clumpiness in the initial conditions can survive. Similarly to the low-density cool fractal above, from an observation of the distribution in Fig.~\ref{fig10} at 2~Myr it would be essentially impossible to reconstruct the initial conditions, and it would be perfectly reasonable to consider this a much more extended star formation event.

\subsubsection{Summary for 30~pc from the GC}

If clusters form well- or just-within their nominal tidal radii they can usually survive for 2~Myr (the age of the Arches) at 30~pc from the GC.  Any cluster which overflows the nominal 2~pc tidal radius is destroyed (if it initially overflows it, or if a dynamical bounce causes this to happen later).

This sets a minimum initial density for an Arches-like cluster forming at 30~pc from the GC of $\sim 600$ M$_\odot$~pc$^{-3}$ irrespective of the initial spatial structure.

At such high densities any initial structure is rapidly erased and so all clusters that survive appear as smooth, spherical distributions by 2~Myr.

All surviving clusters at 30~pc from the GC show tidal features, although they might not be observable if only looking at stars $>2.5$ M$_\odot$.  

Even when no mass segregation was initially present it occurs rapidly in high-density initial conditions due to the short dynamical timescales at these densities. This could cause a difference in the half-mass radius as measure from stars $>2.5$ M$_\odot$ to be 10--20 per cent lower than the true half-mass radius.

{\em Any cluster that has survived to 2~Myr in the strong tidal field at 30~pc from the GC will appear Plummer-like and mass segregated, no matter what its initial conditions were.}

%%%%%%%%%%%%%%%%%%%%%%%%%%%%%%%%%%%%%%%%%%%%%%%%%%%%%%%%%%%%%%%%%%%
\subsection{Clusters at 100 pc from the GC}

\begin{figure*}
	\centering
	\includegraphics[angle=90,width=0.8\linewidth]{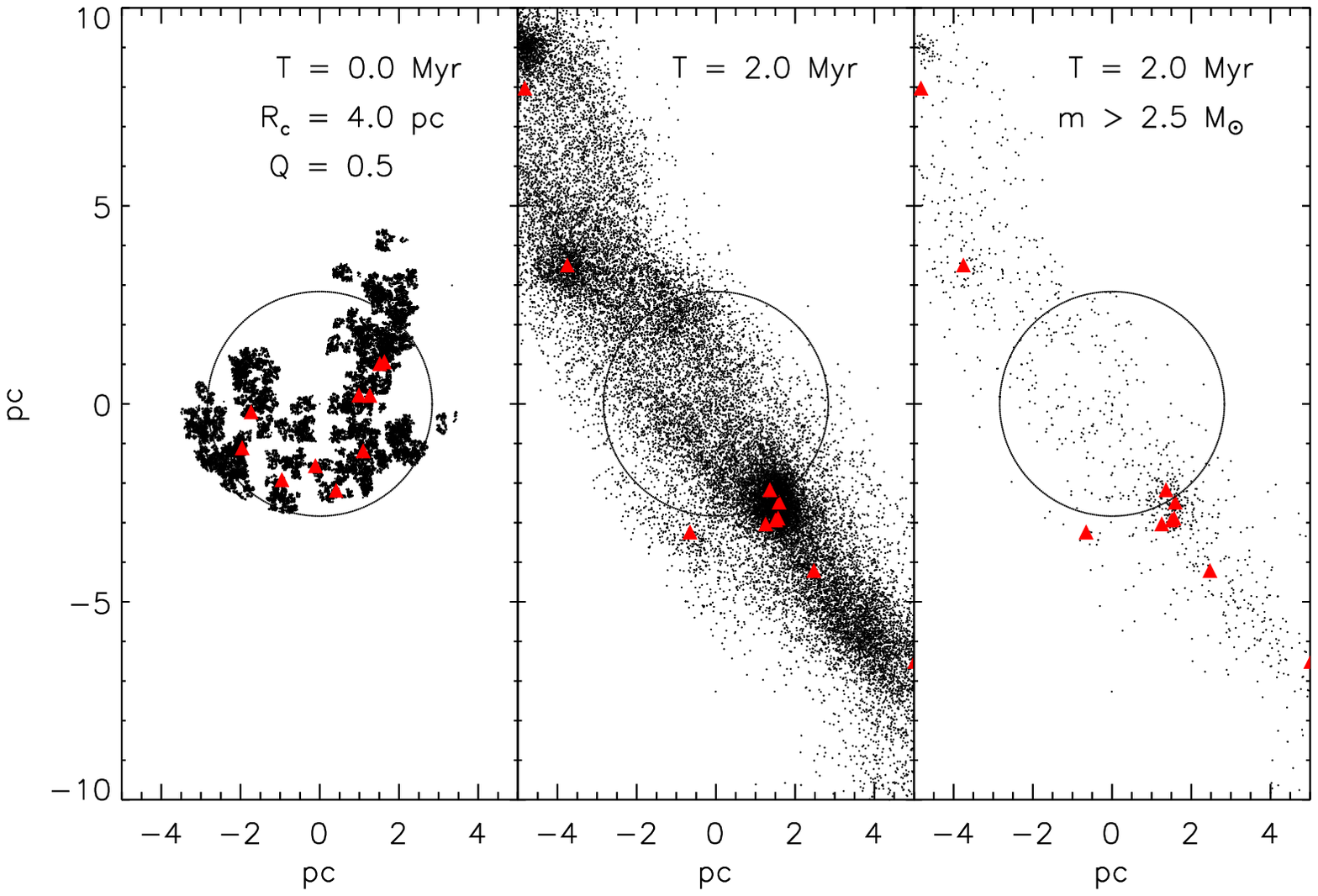}
	\caption{The evolution of an $R_{\rm c} = 4$~pc, tepid ($Q=0.5$) fractal cluster at 100~pc from the GC. The black dots are stars and the red triangles the 10 most massive stars in the cluster. The black circles are the nominal initial tidal radius ($R_{\rm t}$). The left and middle panels show all stars at 0 and 2~Myr, and the right figure shows only stars more massive than $\sim$ 2.5~M$_{\odot}$ at 2~Myr.}
\label{Fig14}
\end{figure*}

\begin{figure*}
	\centering
	\includegraphics[angle=90,width=0.8\linewidth]{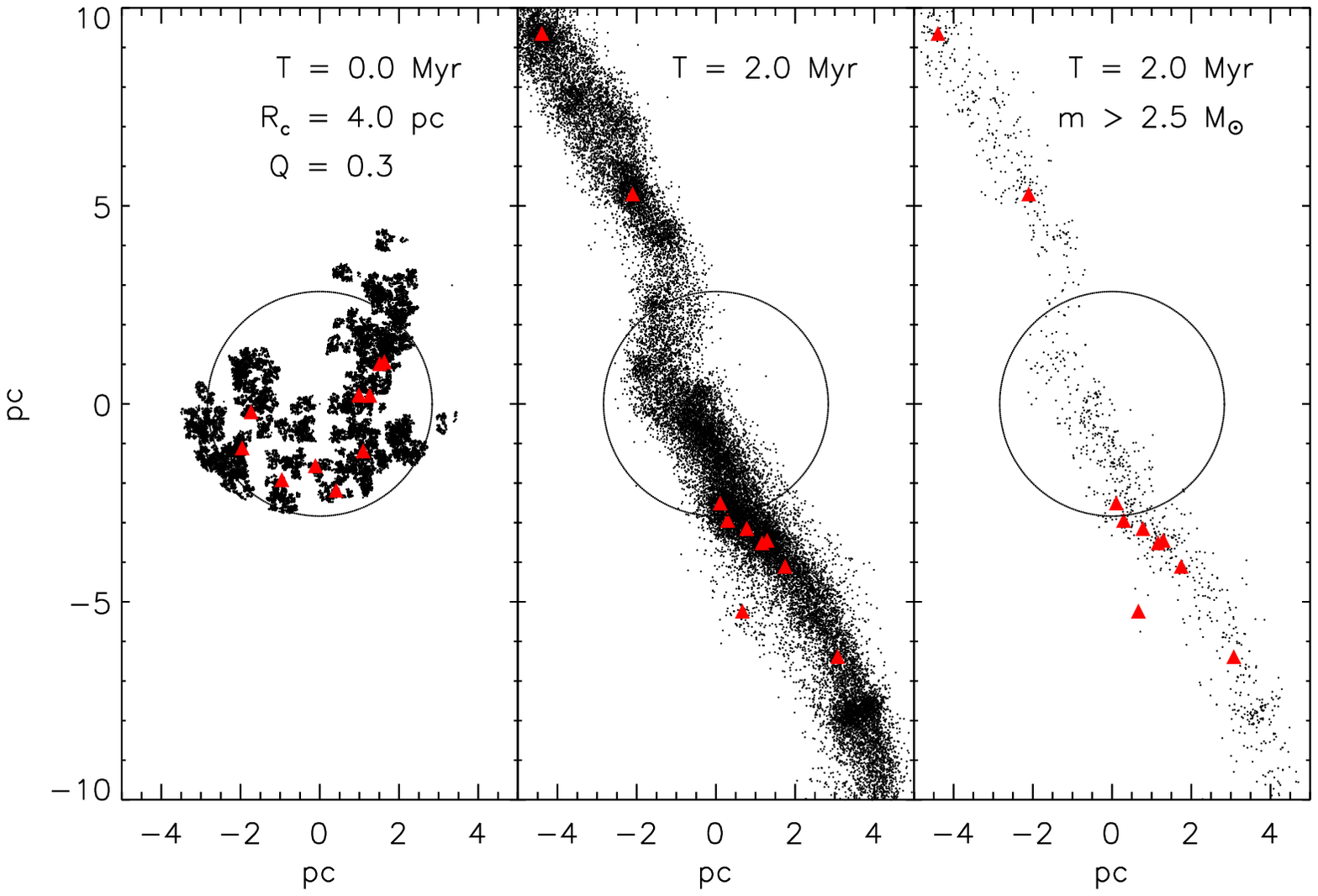}
	\caption{The evolution of an $R_{\rm c} = 4$~pc, cool ($Q=0.3$) fractal cluster at 100~pc from the GC. The black dots are stars and the red triangles the 10 most massive stars in the cluster. The black circles are the nominal initial tidal radius ($R_{\rm t}$). The left and middle panels show all stars at 0 and 2~Myr, and the right figure shows only stars more massive than $\sim$ 2.5~M$_{\odot}$ at 2~Myr.}
\label{Fig15}
\end{figure*}

In this section, we examine the dynamical evolution of star clusters at 100~pc from the GC. The projected separation of the Arches from the GC is $\sim 30$~pc meaning that this is a lower limit on the 3D separation. It is not unreasonable to think that the Arches is at a true distance of, say, 100~pc from the GC \citep{Stolte et al.2008,Habibi et al.2014} which would mean that it is in a significantly weaker tidal field than at 30~pc (although still `strong' by any usual definition of tidal field strength).

At 30~pc from the GC the nominal tidal radius of a $2 \times 10^4$ M$_\odot$ cluster is $\sim 2$~pc, but at 100~pc this increases to $\sim 3$~pc (see Eq.~\ref{eq2.5}). When just filling this larger tidal radius, the density would therefore be a factor of $\sim 3$ times lower.

In Sec.~\ref{section3.1} (at 30~pc from the GC) we found that clusters must initially be within their tidal radii in order to survive as a single cluster. In this section we find that this is still true at 100~pc from the GC with some small caveats. Note that when the results are simply scaled-up versions of those at 30~pc we will not illustrate them.

{\bf Well-contained initial conditions at 100~pc.}
Initial distributions (Plummer and fractal) are well-contained within the nominal tidal radius at 100~pc with $R_{\rm c} = 2$~pc.  These are 8 times lower densities than at 30~pc, but their evolution is very similar. They are able to survive as distinct bound entities for 2~Myr, and fractal initial conditions erase their substructure. As in the 30~pc case, both Plummer and fractal initial conditions appear very similar to each other by 2~Myr. Due to the lower densities, dynamical evolution is slightly slower so they mass segregate to a slightly lesser degree than for well-contained clusters at 30~pc (although if they have the same initial densities, i.e. are very well-contained, they evolve in almost exactly the same way as at 30~pc). This is unsurprising.

{\bf Just-contained initial conditions at 100~pc.} When the initial distributions are just contained within the nominal tidal radius of 3~pc (at 100~pc from the GC), the evolution is somewhat different to when they are just contained within the 2~pc nominal tidal radius (at 30~pc from the GC).

In Fig.~\ref{fig12} we show the evolution of a just-contained virialised Plummer sphere at 100~pc from the GC. This should be compared with Fig.~\ref{fig06} for a just-contained virialised Plummer sphere at 30~pc. The first panels of Figs.~\ref{fig12} and~\ref{fig06} are very similar -- the only difference is that to be just-contained within the nominal tidal radius the Plummer sphere at 100~pc is 1.5 times larger. But at 2~Myr in the second panel, whilst the cluster is being destroyed for the same reason as at 30~pc, the destruction is on a longer timescale due to the lower tidal field strength.  Therefore, the stars have had a chance to interact with each other as well as feel the tidal field.  Interestingly, the massive stars (red triangles) are much less dispersed than the `background' and an observer may still see this as a `cluster-like' object (or even `association-like'), especially when observing only stars $>2.5$ M$_\odot$ as in the third panel.  Therefore a just-contained Plummer sphere is more distinct at 100~pc than at 30~pc (although still is not surviving in any real sense).

For a just-contained tepid fractal at 100~pc as shown in Fig.~\ref{fig13} is interestingly less able to survive as a single entity than its equivalent at 30~pc. The larger just-contained fractal distribution at 100~pc takes longer to relax and erase substructure compared to the equivalent cluster at 30~pc (as shown in Fig.~\ref{fig07}).  This is due to the longer dynamical timescales at the lower density.  In particular, a significant `subcluster' has detached from the main cluster (to the upper left). The mass of the main cluster is $\sim$ 13000~M$_{\odot}$ (with $\sim$ 5700~M$_{\odot}$ in stars > 2.5~M$_{\odot}$), and the mass of the `subcluster' is $\sim$ 1100~M$_{\odot}$ (with $\sim$ 500~M$_{\odot}$ in stars > 2.5~M$_{\odot}$). Any hypothetical observers viewing Fig.~\ref{fig13} would fin it extremely difficult to say if the subcluster at the top left had formed with the main cluster and been `detached', or if it had formed as a separate entity. (A similar subcluster is detached from the 30~pc case, but this is rapidly destroyed by the stronger tidal field).

{\bf Overflowing initial conditions at 100~pc.}  As at 30~pc, overflowing Plummer spheres are completely destroyed and spread along the orbit.  Overflowing fractals do not survive as a single entity, but significant structure and subclusters can remain.  In Fig.~\ref{Fig14} we show the evolution of an overflowing fractal with total radius $R_{\rm c} = 4$~pc (with a nominal tidal radius of 3~pc at 100~pc from the GC).  A significant cluster still survives after 2~Myr (located at roughly 2~pc, -2~pc) which contains 5 of the most massive stars. This is a more extensive surviving subcluster than in Fig.~\ref{fig10} (the equivalent at 30~pc from the GC) as the longer dynamical timescales for the relaxation of the fractal are more than balanced by the weaker tidal field.

In Fig.~\ref{Fig15} we illustrate a cool, overflowing fractal ($Q=0.3$ and $R_{\rm c} = 4$~pc; i.e. the same as that in Fig.~\ref{Fig14} but dynamically cooler).  Again, this fails to survive as a single entity, but there are considerable surviving overdensities, and a `string' slightly below the origin containing 6 of the 10 most massive stars within 2~pc of each-other.  Note that these are all stars with masses $> 50$ M$_\odot$ and so this feature would be very obvious.  As with many features in strong tidal fields, this `filament of massive star formation' (which would be a seemingly sensible interpretation) is not `real'.

\subsubsection{Summary for 100~pc from the GC}

Initial conditions that are well contained within the nominal tidal radius at 100~pc from the GC survive, and mass segregate, and evolve in a very similar way to those at 30~pc from the GC, but can do so in a somewhat longer timescale due to the (potentially) lower density.  {\em Again we find that to survive as a single entity the initial conditions must be contained within the tidal radius, setting a minimum initial density for the Arches if it is (or formed at) 100~pc from the GC of 200~M$_\odot$~pc$^{-3}$.}

Initial conditions that are just-contained within the 3~pc nominal tidal radius at 100~pc from the GC evolve in a slightly subtle way.  The weaker tidal field affects them less (e.g. the initially Plummer distribution is destroyed more slowly), but the lower density means they can evolve internally less (hence fractals can have subclusters removed by the tidal field before they are erased).

Any initial conditions that are able to survive look Plummer-like and tend to be mass segregated at 100~pc from the GC, they can be very similar to those at 30~pc from the GC, and potentially slightly larger (by a factor of roughly 1.5 at most).

\section{Initial conditions and gas}

A key result from these simulations is that for a purely $N$-body star cluster to survive in a strong tidal field it must be initially well-contained within its nominal tidal radius.  Distributions that either start or are able to expand/bounce beyond their nominal tidal radii are rapidly shredded by tidal forces.

Due to computational limitations our simulations are just of the stellar distribution and ignore any gas left-over after star formation.  One would not expect star forming clouds to convert gas to stars at 100 per cent efficiency, and so a (significant) gas potential may remain.  The removal of gas from young clusters can have a significant effect on the stellar component causing it to expand, and even unbind the cluster \citep{Goodwin&Bastian2006, Baumgardt&Kroupa2007}.  However, what effect gas loss has can depend very strongly on the density and velocity distributions of the stars within the gas potential meaning that it is difficult to simply link star formation efficiency to the possible effects of gas loss \citep{Verschueren&David1989, Goodwin2009, Farias et al.2015, Lee&Goodwin2016}.

The effect of loosing whatever gas is left over after star formation will be to cause the stellar distribution to expand to some extent.  Therefore, the minimum densities at which the stellar distributions can have in order to survive are lower limits on the initial densities at which the stars could form, e.g. at 100~pc from the GC the initial stellar distribution must have a density of $> 200$ M$_\odot$~pc$^{-3}$ {\em after any effect of gas expulsion} (whatever that may have been).

\section{Conclusion}

We investigate the early evolution of $\sim 2 \times 10^4$ M$_\odot$ Plummer and fractal initial conditions in a strong tidal fields at 30 and 100~pc from the GC where the nominal tidal radii are $\sim 2$~pc and $\sim 3$~pc respectively. 

We perform $N$-body simulations using {\sc nbody6} \citep{Aarseth1999} with full tidal fields from \citet{Kim et al.2000}. We start our stellar initial conditions either well-contained within the tidal radius, just-filling the tidal radius, or overflowing the tidal radius at both 30 and 100~pc from the GC.  We evolve the clusters for 2~Myr and then compare the final state with each-other and the Arches cluster.

{\em Both Plummer sphere and fractal stellar initial conditions that are well-contained within the nominal tidal radii survive for 2~Myr as distinct bound clusters.} Fractal initial conditions rapidly relax and erase their substructure and become spherical and Plummer-like, and both Plummer spheres and fractal initial conditions are able to dynamically mass segregate within 2~Myr. Both Plummer sphere and fractal initial conditions give rise to tidal tails of low-mass stars that would be difficult to observe at the GC. It is essentially impossible to determine the initial conditions from the state at 2~Myr as both fractals and Plummer spheres produce round, virialised, mass segregated final clusters.

{\em If the stellar initial conditions completely fill the tidal radius then if a single significant cluster survives depends somewhat on the initial conditions.}  Plummer spheres are destroyed, and somewhat unexpectedly, tepid (virial ratio 0.5) fractals survive, but cool (virial ratio 0.3) fractals are destroyed due to a `bounce'.

{\em Stellar initial conditions that overflow the nominal tidal radius are destroyed.}  But if the initial conditions are fractal then significant `subclusters' can survive.  It would be extremely difficult to disentangle the initial conditions, and a rapidly shredded localised star formation event would rapidly look to be a more extended event.

{\em If a single significant cluster survives then after 2~Myr it appears as a mass segregated Plummer-like object irrespective of the initial conditions.}  This is because clumpy initial conditions are dynamically erased, and the high densities cause dynamical mass segregation.  There are subtle signatures in the degree of mass segregation and density profiles to the initial conditions, but it is doubtful they could ever be observed in enough detail to be useful.

The Arches cluster is a 2~Myr old, spherical, virialised, mass-segregated cluster close to the GC. In order to appear like this now, it could have formed as either a smooth Plummer-like, or a clumpy, fractal-like distribution. However, its initial radius must have been $< 2$~pc if it formed at 30~pc from the GC, or $< 3$~pc if it formed at 100~pc from the GC. This sets a lower limit on the formation density of the Arches of $> 600$ M$_\odot$~pc$^{-3}$ at 30~pc, or $> 200$ M$_\odot$~pc$^{-3}$ at 100~pc.

\section*{Acknowledgements}
This work was supported by the National Research Foundation grant funded by the Ministry of Science and ICT of Korea (NRF-2014R1A2A1A11052367). This work was also supported by the BK21 plus program through the National Research Foundation (NRF) funded by the Ministry of Education of Korea. S.-M.P. is deeply thankful for support when visiting the University of Sheffield to work on this paper.

%%%%%%%%%%%%%%%%%%%%%%%%%%%%%%%%%%%%%%%%%%%%%%%%%%



%\bibliographystyle{mnras}


%%%%%%%%%%%%%%%%%%%%%%%%%%%%%%%%%%%%%%%%%%%%%%%%%%

%%%%%%%%%%%%%%%%% APPENDICES %%%%%%%%%%%%%%%%%%%%%

%%%%%%%%%%%%%%%%%%%%%%%%%%%%%%%%%%%%%%%%%%%%%%%%%%

% Don't change these lines
\bsp	% typesetting comment
\label{lastpage}
\end{document}